\documentclass[a4paper,11pt]{article}
\usepackage{amssymb}
\usepackage{amsmath}
\usepackage[dvips]{graphicx}
\usepackage{amsthm}
\usepackage{amscd}
\usepackage{latexsym}
\usepackage{mathrsfs}
\begin{document}
\centerline{\large\bf Asymptotic expansion for cycles in homology
classes for graphs}
\vskip.6cm

\centerline{\bf Dongsheng Liu\footnote{This work was done at
department of mathematics, university of Manchester.}}

\vskip.3cm \centerline{Department of Physics, Lancaster University
Lancaster, LA$1$ $4$YB, UK}

 \centerline{e-mail:
d.liu@lancaster.ac.uk}

 \vskip.7cm
 \noindent
{\bf Abstract}.\quad
  In this paper
we give an   asymptotic expansion including error terms for the
number of cycles in homology classes for connected graphs. Mainly,
we obtain formulae about the coefficients of error terms which
depend on the homology classes and give two examples of how to
calculate the coefficient of first error term.

 \vskip.4cm
\emph{$2000$ Mathematics Subject Classification.} $37$C$27$,
$37$C$30$, $37$D$05$.
 \vskip.6cm

{\bf 1. Introduction.} \ To estimate the number of closed orbits
for certain flows has been studied
 by many authors such as [$\boldsymbol{2}$], [$\boldsymbol{4}$] and
 [$\boldsymbol{9}$]. The error terms of asymptotic expansion were not
 known until  the works of Dolgopyat on  Anosov flows, where he obtained strong results on the contractivity of transfer operator.
  These results led Anantharaman [$\boldsymbol{1}$]
 , Pollicott and Sharp [$\boldsymbol{10}$] and Liu [$\boldsymbol{5}$] to find full expansions of expression for
 the number of closed orbits for Anosov flows. The key to these
 methods lies in reduction of  calculating closed orbits of an  Anosov flow to calculating closed
 orbits of a suspended flow or to calculating  periodic points of a subshift of finite type [$\boldsymbol{8}$].
 \vskip.3cm

 This strategy led us to consider the number of cycles of a connected graph in this article.

\vskip.3cm
 A graph
$G$ is defined to be a pair $(V,E)$, where $V$ is a set
$\{V_1,V_2,\ldots,V_n\}$ of elements called vertices, and $E$ is a
family $(e_1,e_2,\ldots,e_m)$ of (undirected) edges joining
elements of $V$. There may be more than one edge joining the two
vertices. If  a vertex is joined  to itself by a edge, we call
this edge a loop. We will only consider the connected finite
graphs in this article.

It is convenient to speak of graph in which each edge has an orientation attached to it. In this case, we call the graph an oriented graph. We can associate to an undirected graph $G$ with $n$ vertices and $m$ edges ,  an oriented graph $G_o$ with $n$ vertices and $2m$ edges. An oriented graph $G_o$ is a pair $(V,\mathbb{E})$, where $\mathbb{E}$ is a set of ordered pairs of  elements of $V$. For $e\in\mathbb{E}$, we denote by $I(e)$ the initial endpoint of $e$ and $T(e)$ the terminal endpoint of $e$.

We label the edges of oriented graph $G_o$ by $1,2,\ldots, 2m$. For example, Figure $1.1$ is a undirected graph with $3$ vertices and $4$ edges, Figure $1.2$ is the corresponding oriented graph to Figure $1.1$.
\begin{figure}[!ht]
\label{fig1.1}
\begin{center}
\includegraphics[width=11cm]{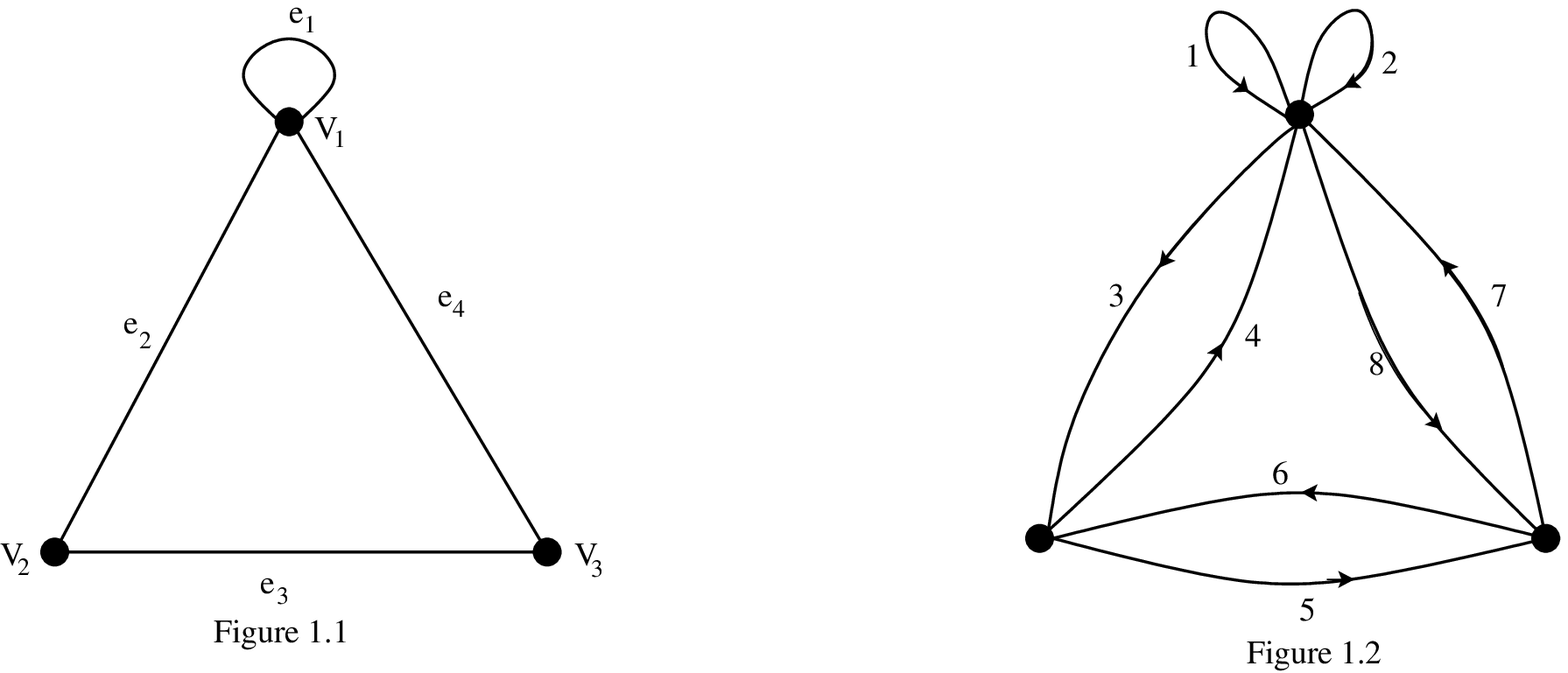}
\end{center}
\end{figure}

A chain is a sequence $u=(u_1,u_2,\ldots,u_q)$ of edges of $G_o$ such that each edge in the sequence has one endpoint in common with its predecessor in the sequence and its other endpoint in common with its successor in the sequence, i.e., $T(u_i)=I(u_{i+1})$, $i=1,2,\ldots,q-1.$

A cycle $\gamma$ is a chain such that the two endpoints of the chain are the same vertex, i.e., a chain $(u_1,u_2,\ldots,u_q)$ is a cycle if $T(u_q)=I(u_1)$. The edge length of a cycle $\gamma$ is defined by the number of edges in $\gamma$. We say a cycle $\gamma=(u_1\ldots,u_q)$ has backtracking if $u_i=-u_{i+1}$ for some $i$, $1\leq i\leq q-1$, where $-u_{i+1}$ is the reverse of $u_{i+1}$.

We assign a length to each edge and denote  the length of $e_i$ by $l(e_i)$. For the corresponding oriented graph, we have $l(e_i)=l(-e_i)$. The length of a chain $(u_1,u_2,\ldots, u_n)$ is $l(u_1)+l(u_2)+\cdots+l(u_n)$.

We denote by $H_1(G,\mathbb{Z})$ the homology group of $G$. For
convenience, we assume that  $H_1(G,\mathbb{Z})=\mathbb{Z}^b$.
Otherwise, we can write  $H_1(G,\mathbb{Z})=\mathbb{Z}^b\oplus
H$. Since the torsion subgroup $H$ is finite, the results then will
only differ by a multiplicative constant.

\vskip.2cm
 Let $\Gamma$ be the set of cycles in graph $G$. For
$\gamma\in \Gamma$ we denote by $[\gamma]$ the homology class in
$H_1(G,\mathbb{Z})$. Let $l(\gamma)$ be the length of $\gamma$.

 For $\alpha\in H_1(G,\mathbb{Z})$, let
$$\pi(T,\alpha)=\#\{\gamma\in\Gamma, l(\gamma)\leq T,  [\gamma]=\alpha\}.$$

We will give the asymptotic formulae  for $\pi(T,\alpha)$  which
is similar to the case of homologically full transitive Anosov
flow [$\boldsymbol{5}$]. But we will concentrate on how to
calculate the first error term for special cases in this paper.

\vskip.3cm

 We briefly outline the contents of this article. In section $2$, we
explain how, through the use of symbolic dynamics, the counting
problem for cycles can be reduced to one for periodic points for a
subshift of finite type. In section $4$, we introduce a function
$Z(s,v)$ and derive some important properties of its analytic
extension which be used to obtain the formula for distribution of
cycles including error terms, that is Theorem $1$. We specify the
coefficient for the first error term in this section. In the last
two sections we will give two examples for how to calculate the
coefficient of the first error term, where we use two different
methods. Since the calculating of coefficient of first error term
involve the derivatives of a function $\beta (u)$ which be
introduced in section $3$, we will give some formulae of
derivatives of $\beta (u)$ in this section.

 \vskip.6cm
 {\bf 2. Symbolic dynamics.} \  For a graph
$G$ with $m$ edges there exists a $2m\times 2m$ matrix $A_G$ with
zero-one entries associated with the corresponding oriented graph
$G_o$. The matrix $A_G$ can be defined by following. For $1\leq
i,j \leq 2m$, if the terminal endpoint of edge $i$ is equal to the
initial endpoint of edge $j$ then $A(i,j)=1$, otherwise
$A(i,j)=0$. For example, the matrix associated with Figure $1.2$
is

\begin{displaymath}
A_G=
\left(\begin{array}{cccccccc}
1 & 1 & 1 & 0 & 0 &0 &0&1\\
1 &1 &1& 0& 0&0& 0&1\\
0&0&0&1&1&0&0&0\\
1&1&1&0&0&0&0&1\\
0&0&0&0&0&1&1&0\\
0&0&0&1&1&0&0&0\\
1&1&1&0&0&0&0&1\\
0&0&0&0&0&1&1&0\\
\end{array}
\right)
\end{displaymath}
\vskip.4cm Let $A=(a_{ij})$ be a $k\times k$ matrix, we say $A$ is
non-negative if $a_{ij}\ge 0$ for all $i,j$. Such a matrix is
called irreducible if for any pair $i,j$ there is some $n$ such
that $a_{i,j}^{(n)}>0$ where $a_{ij}^{(n)}$ is $(i,j)$-th element
of $A^n$, i.e.,  $a_{ij}^{(n)}=A^n_{ij}$. The matrix $A$ is
aperiodic if there exists $n>0$ such that $a_{ij}^{(n)}>0$ for all
$i,j$.

It is easy to see that the graph $G$ is connected if and only if
associated $A_G$ is irreducible. So if $A_G$ is aperiodic then $G$ is
connected. But if $G$ is connected, $A_G$ may be not aperiodic, for
example, a bipartite graph is connected but the associated matrix is
not aperiodic. Where we say a graph $G$ is  bipartite if its vertex set can be partitioned into two classes such that no two adjacent vertices belong to the same class. A graph is bipartite if and only if it possesses no cycles of odd edge length.

\noindent However, It is easy to prove that
\newtheorem {le40}{Lemma}
\begin{le40}
If $G$ is connected and it is not bipartite, then associated matrix $A_G$ is aperiodic.
\end{le40}

\vskip.3cm We will only consider connected graph $G$ whose
corresponding matrix $A_G$ is aperiodic.

 We define $\Sigma_{A} $ by
$$\Sigma_A=\left\{x\in\prod_{0}^{\infty}\{1,2,\ldots, 2m\}: A_{G}(x_i,x_{i+1})=1, \forall i\in \mathbb{Z}^+\right\}.$$
The subshift of finite type $\sigma: \Sigma_{A}\to \Sigma_{A}$ is defined by
subshift$$(\sigma x)_i=x_{i+1}.$$
We define $r: \Sigma_{A}\to \mathbb{R}^+$ by
$r(x)=l(x_0),$ then
$$l(x_0,x_1,\ldots,x_{n-1})=r(x)+r(\sigma x)+\cdots+r(\sigma^{n-1}x)=:r^n(x).$$
There is a one-one correspondence between closed orbits $\{x,\sigma x,\ldots,\sigma^{n-1} x\}$ for $\sigma: \Sigma_A\to \Sigma_A$ and cycles of the graph $G$. The least length of corresponding cycle is $r^n(x)$.

There exists $f=(f_1,f_2,\ldots,f_b) : \Sigma_A \to \mathbb{R}^b$
such that for $\gamma\in \Gamma$, $[\gamma]=f^n(x)$ for some $n,
x$ with $\sigma^n x=x$. We can even make $f(x)$ just depend on one
co-ordinate, i.e., $f(x)=f(x_0)$  [$\boldsymbol{8}$].

\vskip.2cm
\emph{Remark}:\\
If $A_{G}$ is connected but not aperiodic then it is a bipartite
graph. In this case, we can decompose $\Sigma_A$ by
$\Sigma_A=\Sigma_0\bigcup \Sigma_1$ satisfies
$$\sigma: \Sigma_0\longrightarrow \Sigma_1, \qquad \Sigma_1\longrightarrow \Sigma_0.$$
So $\sigma^2: \Sigma_A\to \Sigma_A$ satisfies $\sigma^2:
\Sigma_0\to \Sigma_0, \qquad \Sigma_1\to \Sigma_1$. we define
$$R(x):=r^2(x)=r(x)+r(\sigma x)$$
and
$$F(x):=f^2(x)=f(x)+f(\sigma x).$$
There is one-one correspondence between closed orbits
$\{x,(\sigma^2) x,\ldots,(\sigma^2)^{n-1} x\}$ for $\sigma^2:
\Sigma_0\to \Sigma_0$ or $(\Sigma_1\to \Sigma_1)$ and cycles of
the graph $G$. The least length of corresponding cycle is
$R^n(x)$, which is same as $A_G$ is aperiodic.

\vskip.2cm
In order to obtain a positive result we shall only consider the graphs satisfying the following conditions [$\boldsymbol{10}$].\\

\noindent (A) \emph{Weak-Mixing}. The closed subgroup of $\mathbb{R}$ generated by $\{l(\gamma)\} (\gamma\in\Gamma)$ is $\mathbb{R}$.

\noindent (B) \emph{Approximability Condition.} There exist three
cycles $\gamma_1$, $\gamma_2$ and $\gamma_3$ with least lengths
$l(\gamma_1)$, $l(\gamma_2)$ and $l(\gamma_3)$, respectively, such
that
$$\xi=\frac{l(\gamma_1)-\l(\gamma_2)}{l(\gamma_2)-l(\gamma_3)}$$
is badly approximable, i,e., there exists $\alpha>0$ and $C>0$
such that we have $|\xi-\frac{p}{q}|\ge \frac{C}{q^{\alpha}}$, for
all $p,q \in \mathbb{Z} (q>0).$

The set of $\xi$ satisfying this condition is a large set. For
example, it is a set of full measure in the real line. Moreover,
its complement has Hausdorff dimension zero.

 \vskip.6cm
 {\bf 3.
 Derivatives of function $\beta (u)$.} \  In this section, we first briefly review the pressure function
then calculate the derivatives of
 associated function $\beta (u)$.  The pressure function $P: C(\Sigma_A)\to \mathbb{R}$ is
defined by
$$P(g)=\sup_{m\in M_{\sigma}}^{}\{h_m(\sigma)+\int gdm\},$$
where $M_{\sigma}$ is the set of $\sigma$-invariant probability measures and $h_{m}(\sigma)$ is the entropy of $\sigma$ with respect to $m$. Let $h$ be the unique number such that $P(-hr)=0$. There is no loss in generality in assuming $\int fd\mu_{-hr}=0$, where $\mu_{-hr}$ is the equilibrium state of $-hr$.

For $u\in \mathbb{R}^b$, the function $\beta (u): \mathbb{R}^b\to \mathbb{R}$ is defined by
\begin{equation}
\label{4.3.1}
P(-\beta (u)r+\langle u,f\rangle)=0
\end{equation}
Then $\beta (u)$ is an analytic function on $\mathbb{R}^b$ and $\beta (u) $ is strictly convex in each $u_i$, $i=1,2,\ldots, b$, where $\langle u,f\rangle=\sum_{i=1}^b u_if_i$. Now we  can extend $\beta  (u)$ to complex values of the argument. For all $u\in\mathbb{R}^b$, $P(-\beta (u)r+\langle u,f\rangle)=0$ and $P(-sr+\langle u+iv,f\rangle)$ is analytic for $(s,u+iv)$ in a neighbourhood of $\mathbb{R}\times \mathbb{R}^b$ in $\mathbb{C}\times\mathbb{C}^b$. Since
$$\left[\frac{\partial P(-sr+\langle u,f\rangle)}{\partial s}\right]_{s=s_0}=-\int rd\mu_{-s_0r+\langle u,f\rangle}\ne 0,$$
by the implicit function theorem, $\beta (u)$ can extend to an analytic function on a neighbourhod of $\mathbb{R}^b$ in $\mathbb{C}^b$ by the equation
$$P(-\beta (u+iv)r+\langle u+iv,f\rangle)=0.$$
We have $\beta (0)=h$,  since $P(-hr)=0$.

When  estimating $\pi (T,\alpha)$, the formulae which arise involve derivatives of the function $ \beta (u)$. In this section, we shall calculate these derivatives up to the fourth order. Partial differentiating  (\ref{4.3.1}) with respect to $u_i$ yields
\begin{equation}
\label{4.3.2}
\frac{\partial P}{\partial \beta}\frac{\partial \beta}{\partial u_i}+\frac{\partial P}{\partial u_i}=0.
\end{equation}
Since
$$\left[\frac{\partial P(-\beta (u)r)}{\partial \beta}\right]_{\beta=\beta (0)=h}=-\int rd\mu_{-hr},$$
and
$$\left[\frac{\partial P(-hr+<u,f>)}{\partial u_i}\right]_{u=0}=\int f_id\mu_{-hr},$$
we have
$$\frac{\partial \beta (0)}{\partial u_i}=\frac{\int f_id\mu_{-hr}}{\int rd\mu_{-hr}}=0.$$
For obtaining expression of $\nabla \beta ^2(0)$, partial differentiate (\ref{4.3.2}) with respect to $u_j$,
and note  $\nabla \beta (0)=0$ we have
$$
\frac{\partial ^2 \beta (0)}{\partial u_i\partial u_j}=
\frac{1}{\int rd\mu_{-hr}}\left[\frac{\partial ^2P(-hr+<u,f>)}{\partial u_i\partial u_j}\right]_{u=0}.$$
There is another expression for $\partial^2 \beta(0)/\partial u_i\partial u_j$, that is
$$ \frac{\partial^2 \beta(0)}{\partial u_i\partial u_j}=
 \frac{1}{\int rd\mu_{-hr}}\lim_{n\to\infty}^{}\frac{1}{n}\int f_i^nf_j^nd\mu_{-hr}.$$
We refer to [$\boldsymbol{4}$] for this formula.\\
\vskip.3cm \noindent The third and fourth order derivatives with
respect to $u_i$ are following.
\begin{eqnarray*}
&&\frac{\partial ^3 \beta (0)}{\partial u_i\partial u_j \partial u_m}=\frac{1}{\int rd\mu_{-hr}}\left\{\left[\frac{\partial ^3P(-hr+<u,f>)}{\partial u_i\partial u_j\partial u_m}\right]_{u=0}\right.\nonumber\\
&&+\left.\left[\left(\frac{\partial^2P}{\partial \beta\partial u_i}\frac{\partial^2 \beta}{\partial u_j\partial u_m}+\frac{\partial^2P}{\partial \beta\partial u_j}\frac{\partial^2\beta}{\partial u_i\partial u_m}+\frac{\partial^2 P}{\partial \beta\partial u_m}\frac{\partial^2\beta}{\partial u_i\partial u_j}\right)(-\beta r+<u,f>)\right]_{\beta =h,u=0}\right\} \nonumber.
\end{eqnarray*}

\begin{eqnarray*}
&&\frac{\partial ^4 \beta (0)}{\partial u_i\partial u_j \partial u_m\partial u_n}=\frac{1}{\int rd\mu_{-hr}}\left\{\left[\frac{\partial ^4P(-hr+<u,f>)}{\partial u_i\partial u_j\partial u_m\partial u_n}\right]_{u=0}\right.\\
&&+\left[ \frac{\partial^2P}{\partial \beta^2}\left(\frac{\partial^2\beta}{\partial u_i\partial u_j}\frac{\partial^2\beta}{\partial u_m\partial u_n}+\frac{\partial^2\beta}{\partial u_i\partial u_m}\frac{\partial^2\beta}{\partial u_j\partial u_n}+\frac{\partial^2\beta}{\partial u_i\partial u_n}\frac{\partial ^2\beta}{\partial u_j\partial u_m} \right)\right]\\
&&+\left(\underbrace{\frac{\partial^3P}{\partial \beta\partial u_i\partial u_j}\frac{\partial ^2\beta}{\partial u_m\partial u_n}+\cdots+\frac{\partial^3P}{\partial \beta\partial u_m\partial u_n}\frac{\partial \beta}{\partial u_i\partial u_j}}_{6\quad items}\right)\\
&&+\left.\left.\left(\underbrace{\frac{\partial^2P}{\partial \beta\partial u_i}\frac{\partial^3\beta}{\partial u_j\partial u_m\partial u_n}+\cdots}_{4\quad items}\right) (-\beta r+<u,f>)\right|_{\beta=h,u=0}\right\}.
\end{eqnarray*}
For $k>4$, $\nabla \beta^k(0)$ is more complicated. But for some
special graph $G$, $\nabla \beta^{k}(0)$ may be easy to calculate.
\vskip.6cm

{\bf 4. Distribution of cycles.} \  Let $g$ be of class
$C^{\infty}$ with compact support. For $\alpha\in
H_1(G,\mathbb{Z})$, we first estimate the auxiliary function
$$\pi_g(T,\alpha)=\sum_{\gamma\in\Gamma, [\gamma]=\alpha}^{}g(l(\gamma)-T).$$

Let $\hat{g}$ be the Fourier transform of $g$, By Fourier's Inverse Transform Formula,

\begin{eqnarray*}
\pi_g(T,\alpha)&=&\sum_{\gamma\in \Gamma}^{}
 \frac{1}{2\pi}\int_{\mathbb{R}}\int_{\mathbb{R}^b/\mathbb{Z}^b}\hat{g}(-i\sigma+t)e^{-(\sigma+it)(l(\gamma)-T)}e^{\langle 2\pi i v,[\gamma]\rangle}e^{-\langle 2\pi i v,\alpha\rangle}dvdt\\
&=& \frac{1}{2\pi}\int_{\mathbb{R}}\int_{\mathbb{R}^b/\mathbb{Z}^b}Z(\sigma+it,v)e^{(\sigma +it)T} \hat{g}(-i\sigma+t)e^{-2\pi i <v,\alpha>}dvdt,
\end{eqnarray*}
where we have defined
$$Z(s,v)=Z(\sigma+it,v)=\sum_{n=1}^{\infty}\frac{1}{n}\sum_{x\in
Fix_n}^{} e^{-(\sigma+it)r^n(x)+2\pi
i<v,f^n(x)>}+A(\sigma+it,v),$$ with $Fix_n=\{x\in
\Sigma_A,\sigma^nx=x\}$ and $A(\sigma +it,v)$ is analytic when
$\sigma>h-\epsilon$ for some $\epsilon>0$.

It is well-known that when $Re s=\sigma>\beta(0)=h$, $Z(s,v)$ is
absolutely convergent. For the behaviour of $Z(s,v)$ in $Res<h$, we can determine the domain of $Z(s,v)$ by studying the norm of the transfer operator $\mathcal{L}_{s,v}$,\\
 which is detailed discussed in [$\boldsymbol{3}$].
Same procedure as that in [$\boldsymbol{5}$] or more originally in
[$\boldsymbol{10}$], we have following proposition.
\newtheorem{pro40}{Proposition}
\begin{pro40}
 Under conditions (A) and (B),  there exist $B>0$, $c>0$, $\epsilon >0$, $\lambda>0$, $\rho >0$ and a open set $V_0$, a neighbourhood of $0$
 in $\mathbb{R}^b/\mathbb{Z}^b$ such that\\
(1) $Z(s,v)$ is analytic in $\{s=\sigma+it: \sigma>h-\frac{c}{|t|^{\rho}}, |t|>B\} $. And in this domain $|Z(s,v)|=O(|t|^{\lambda})$;\\
(2) $Z(s,v)+\log(s-\beta(iv))$ is analytic in $\{(s,v): v\in V_0, \sigma>h-\epsilon, |t|\leq B\}$;\\
(3) $Z(s,v)$ is analytic in $\{(s,v):  v\notin \bar{V_0}, \sigma> h-\epsilon, |t|\leq B\}$.
\end{pro40}

\vskip.3cm
  Using Proposition $1$, we can estimate
$$\pi_g(T,\alpha)= \frac{1}{2\pi}\int_{\mathbb{R}}\int_{\mathbb{R}^b/\mathbb{Z}^b}Z(\sigma+it,v)e^{(\sigma +it)T}
\hat{g}(-i\sigma+t)e^{-2\pi i <v,\alpha>}dvdt.$$ We divide
$\mathbb{R}^b/\mathbb{Z}^b$ into $V_0$ and
$\mathbb{R}^b/\mathbb{Z}^b-V_0$ ($V_0$ is a neighbourhood of $0$
 in $\mathbb{R}^b/\mathbb{Z}^b$ in proposition 1).  For $v\in
\mathbb{R}^b/\mathbb{Z}^b-V_0$, $Z(s,v)$ is analytic in
$\{s=\sigma+it: \sigma>h-\epsilon, |t|<B\}\cup\{s=\sigma+it:
\sigma>h-\frac{c}{|t|^{\rho}},|t|>B\}$. It is easy to estimate the
integral over $\mathbb{R}^b/\mathbb{Z}^b-V_0$. For $v\in V_0$,
using suitable contour integral and Residue Formula we can
transfer the integral over $\sigma>h$ to integral over
$\{\sigma+it: \sigma>h-\frac{c}{|t|^{\rho}},|t|>B\}$. Then
expanding the integral function by Taylor Formula we can estimate
the integral over $V_0$. The details are similar to that for
estimating closed orbits in homology class for Anosov flow
[$\boldsymbol{5}$]. We have

\newtheorem{th41}{Theorem}
\begin{th41} Let $G$ be connected finite undirected graph.Assume that $H_1(G,\mathbb{Z})=\mathbb{Z}^b$. If $g$ is of class $C^{\infty}$ with compact support, there exist $h>0$ such that
\begin{equation}
\label{th}
\pi_g(T,\alpha)=\frac{e^{Th}}{T^{b/2+1}}\left(\sum_{n=0}^N
\frac{c_{n,g}(\alpha)}{T^{n}}+O\left(\frac{1}{T^{N+1}}\right)\right)
\textrm{ as } T\to \infty,
\end{equation}
for all $N\in\mathbb{N}$.
If we write $\alpha=(\alpha_1,\alpha_2,\ldots,\alpha_b)$ then the constants $c_{n,g}(\alpha)=\sum_{i_1+i_2+\cdots +i_b=0}^{2n}c_{i_1i_2\ldots i_b}\alpha_1^{i_1}\alpha_2^{i_2}\cdots \alpha_b^{i_b}$, $c_{i_1\ldots i_b}$ are constants which only depend on the lengths of edges of graph $G$ and $g$.
\end{th41}
We assume that $\rho<1$ in Proposition $1$ and let
$\delta=[\frac{1}{\rho}]-1$, same as that in [$\boldsymbol{5}$]
for homologically full transitive Anosov flow, the error term is
not worse than $O\left(\frac{1}{T^{\delta}}\right)$ when we use
approximation argument to estimate $\pi(T,\alpha)$. We have
following theorem.
\newtheorem{th411}[th41]{Theorem}
\begin{th411} Let $G$ be connected finite undirected graph, $H_1(G,\mathbb{Z})=\mathbb{Z}^b$. There exist $h>0$ and $\delta>0$ such that
$$\pi(T,\alpha)=\frac{e^{Th}}{T^{b/2+1}}\left(c_0+\sum_{n=1}^N \frac{c_{n}(\alpha)}{T^{n}}+O\left(\frac{1}{T^{\delta}}\right)\right) \textrm{ as } T\to \infty$$
for  $N=\delta-1$, where $c_0>0$ is a constant.
If we write $\alpha=(\alpha_1,\alpha_2,\ldots,\alpha_b)$ then the constants $c_{n}(\alpha)=\sum_{i_1+i_2+\cdots +i_b=0}^{2n}c_{i_1i_2\ldots i_b}\alpha_1^{i_1}\alpha_2^{i_2}\cdots \alpha_b^{i_b}$, $c_{i_1\ldots i_b}$ are constants which only depend on the lengths of edges of graph $G$.
\end{th411}

\noindent Analogy to the calculating  closed geodesics in
[$\boldsymbol{6}$], the coefficient $c_{1,g}(\alpha)$ in
(\ref{th}) is following.
$$c_{1,g}(\alpha)=-\sum_{i,j=1}^{b}a_{ij}\alpha_i\alpha_j+\sum_{i=1}^{b}b_i\alpha_i+c_1$$
with
\begin{eqnarray*}
a_{ij}&=&2\pi^2\hat{g}(-ih)\int_{\mathbb{R}^b}e^{-\frac{1}{2}\beta^{''}(0)(v,v)}v_iv_jdv,\\
b_i&=&\int_{\mathbb{R}^b}e^{-\frac{1}{2}\beta^{''}(0)(v,v)}F_1(iv)\cdot (2\pi iv)dv,\\
c_1&=&\int_{\mathbb{R}^b}e^{-\frac{1}{2}\beta^{''}(0)(v,v)}F_2(iv)dv.
\end{eqnarray*}
Where
\begin{eqnarray*}
F_1(iv)&=&\frac{1}{6}\hat{g}(-ih)\beta^{(3)}(0)\cdot(iv)^3+\bar{g}_0^{(1)}(0)\cdot (iv),\\
F_2(iv)&=&\frac{1}{72}\hat{g}(-ih)\left[2(\beta^{(3)}(0)\cdot (iv)^3)^2+3\beta^{(4)}(0)\cdot (iv)^4\right]\\
&+&\frac{1}{6}\bar{g}_0^{(1)}(0)\cdot (iv)
\beta^{(3)}(0)\cdot(iv)^3+\frac{1}{2}\bar{g}_0^{(2)}(0)\cdot(iv)^2+\bar{g}_1^{(0)}(0),
\end{eqnarray*}
with $\bar{g}_j(iv)=\frac{d^j \hat{g}(-i\beta (iv))}{ds^j}$.\\

Since $\beta^{''}(0)$ is positive definite, there exists a linear transformation $v=Mu$ such that $\langle v, \beta^{''}(0)v\rangle=\sum_{k=0}^{b}u_k^2$, where $v=(v_1,v_2,\ldots,v_b)^T$, $u=(u_1,u_2,\ldots,u_b)^T$, $M$ is a  $b\times b$ matrix with $\det M>0$. That is, there exists a matrix $M$ such that $M^T\beta^{''}(0)M=Id$. Hence
\begin{eqnarray*}
a_{ij}&=&2\pi^2\hat{g}(-ih)\int_{\mathbb{R}^b}e^{-\frac{1}{2}\sum_{k=1}^{b}u_k^2}\left(\sum_{l=1}^{b}M_{il}u_l\right)\left(\sum_{m=1}^{b}M_{jm}u_m\right)\det M du\\
&=&2\pi^2\hat{g}(-ih)\det M \sum_{l=1}^{b}M_{il}M_{jl}\int_{\mathbb{R}^b}e^{-\frac{1}{2}\sum_{k=1}^{b}u_k^2}u_l^2du\\
&=&(2\pi)^{\frac{b}{2}+2}\frac{\hat{g}(-ih)}{2}\det M \sum_{l=1}^{b}M_{il}M_{jl}.
\end{eqnarray*}
It is easy to see  $b_i=0$ in here because $\pi_g(T,\alpha)=\pi_g(T,-\alpha)$. The formula for the constant $c$  is still complicated since we need to calculate $\beta^{(3)}(0)$ and $\beta^{(4)}(0)$.

\vskip.2cm

 We  take $g$ close $\chi_{[-\infty,0]}$, then
$\pi_g(T,\alpha)=\pi(T,\alpha)$. Furthermore
$$\hat{g}(-is)=\int_{-\infty}^{0}e^{sy}dy=\frac{1}{s}.$$
Hence $\hat{g}(-ih)=\frac{1}{h}$ and
$\bar{g_0}(-is)=\hat{g}(-is)=\frac{1}{s}$. However
$\bar{g}_0(iv)=\hat{g}(-i\beta (iv))=\frac{1}{\beta(iv)}$ and
 $\bar{g}_1(iv)=-\frac{1}{\beta^2(iv)}$. In this case,
$\bar{g}_0^{(1)}(0)=0$,
$\bar{g}_0^{(2)}(0)=\frac{\nabla^2 \beta (0)}{h^2}$ and
$\bar{g}_1^{(0)}(0)=-\frac{1}{h^2}$.

\noindent So
 $$a_{ij}=\frac{(2\pi)^{\frac{b}{2}+2}}{2h}det M \sum_{l=1}^{b}M_{il}M_{jl}.$$
\newtheorem{th42}[th41]{Theorem}
\begin{th42} Let $G$ be connected finite undirected graph. There exist $h>0$ and $\delta>0$ that that
$$\pi(T,\alpha)=\frac{e^{Th}}{T^{b/2+1}}\left(c_0+\sum_{n=1}^N \frac{c_{n}(\alpha)}{T^{n}}+O\left(\frac{1}{T^{\delta}}\right)\right) \textrm{ as } T\to \infty $$ with
$$c_{1}(\alpha)=-\frac{(2\pi)^{\frac{b}{2}+2}}{2h}det M \sum_{i,j=1}^{b}\sum_{l=1}^{b}M_{il}M_{jl}\alpha_i\alpha_j+c_{1,0},$$
where $M=(M_{ij})$ is a $b\times b$ matrix such that
$(MM^T)^{-1}=\beta^{''}(0)$ and  $c_{1,0}$ is a constant which is independent of $\alpha$.
\end{th42}
\vskip.6cm

{\bf5.  Example $1$.} \  Let us consider simple case, where $G$ is
a graph with one vertex and $k$ edges which form $k$ loops. In
this case,
\begin{displaymath}
A_G=\left(
\begin{array}{ccc}
1&\ldots&1\\
\vdots&&\vdots\\
1&\ldots&1
\end{array}
\right)
\end{displaymath}
Let the lengths of edges be $l_1, l_2,\ldots, l_k$ respectively such that conditions $(A)$ and $(B)$ are satisfied.

We define $r: \Sigma_A\to \mathbb{R}$ by
\begin{displaymath}r(x)=r(x_0)=\left\{\begin{array}{ll}
l_1&\textrm{if $x_0=1$}\\
l_1 &\textrm{if $x_0=2$}\\
\cdots&\cdots\\
l_k &\textrm{if $x_0=2k-1$}\\
l_k &\textrm{if $x_0=2k$ }
\end{array}\right.
\end{displaymath}
The homology group of $G$, $H_1(G,\mathbb{Z})\cong \mathbb{Z}^k$.

$f: \Sigma_A\to \mathbb{Z}^k $ is defined by  $f(x)=f(x_0)=(f_1(x_0),\ldots,f_k(x_0))$ such that
\begin{displaymath}
f_i(x)=f_i(x_0)=\left\{\begin{array}{ll}
1&\textrm{if $x_0=2i-1$}\\
-1 &\textrm{if $x_0=2i$}\\
0 &\textrm{otherwise}
\end{array}\right.
\end{displaymath}
In order to obtain the formula $c_1(\alpha)$, we need to calculate
$\beta^{''}(0)$ and $F_1(iv)$, $F_2(iv)$ which involve
$\beta^{(3)}(0)$ and $\beta^{(4)}$. Next we compute
$\beta^{''}(0)$, $\beta^{(3)}(0)$ and $\beta^{(4)}(0)$.

 Noting $P(-hr)=0$, we have
\begin{equation}
\label{4.6.1}
e^{P(-hr+<u,f>)}=\sum_{l=1}^{2k}e^{-hr(l)+<u,f(l)>}.
\end{equation}
In this case,
$$\int
rd\mu_{-hr}=\sum_{l=1}^{2k}r(l)e^{-hr(l)}=2\sum_{i=1}^{k}l_ie^{-hl_i}.$$
By directly calculattion, we have
$$\left[\frac{\partial P(-hr+<u,f>)}{\partial
u_i}\right]_{u=0}=\sum_{l=1}^{2k}e^{-hr(l)}f_i(l)=e^{-hl_i}[1+(-1)]=0,$$
\begin{displaymath}
\left[\frac{\partial^2P(-hr+<u,f>)}{\partial u_i\partial u_j}\right]_{u=0}=
\left\{\begin{array}{ll}
2e^{-hl_i}&\textrm{if $i=j$}\\
0&\textrm{if $i\ne j$},
\end{array}\right.
\end{displaymath}
$$\left[ \frac{\partial^2P}{\partial \beta\partial u_i}(-\beta r+<u,f>)\right]_{\beta=h,u=0}=0, \forall i,$$
$$\left[\frac{\partial^3P(-hr+<u,f>)}{\partial u_i\partial u_j\partial u_m}\right]_{u=0}=\sum_{l=1}^{2k}f_i(l)f_j(l)f_m(l)e^{-hr(l)}=0.$$
Useing the formulae in section $3$, we have
$$\nabla \beta (0)=0,$$
\begin{displaymath}
\frac{\partial^2\beta(0)}{\partial u_i\partial u_j}=
{\left\{\begin{array}{ll}
\frac{e^{-hl_i}}{\sum_{i=1}^{k}l_ie^{-hl_i}}&\textrm{if $i=j$}\\
0&\textrm{if $i\ne j$},
\end{array}\right.}
\end{displaymath}
$$\frac{\partial^3 \beta (0)}{\partial u_i\partial u_j\partial u_m} = 0.$$

For calculating $\beta^{(4)}(0)$, we also need followings.
\begin{displaymath}
\left[\frac{\partial^4P(-hr+<u,f>)}{\partial u_i\partial u_j\partial u_m\partial u_n}\right]_{u=0}=
\left\{\begin{array}{ll}
-4e^{-h(l_i+l_m)}&\textrm{if $i=j\ne m=n$}\\
-4e^{-h(l_i+l_n)}&\textrm{if $i=m\ne j=n$}\\
-4e^{-h(l_i+l_j)}&\textrm{if $i=n\ne j=m$}\\
2e^{-hl_i}-12e^{-2hl_i}&\textrm{if $i=j=m=n$}\\
0&\textrm{otherwise},
\end{array}\right.
\end{displaymath}

$$\left[\frac{\partial ^2P}{\partial \beta^2}(-\beta r+<u,f>)\right]_{\beta=h,u=0}=2\sum_{i=1}^{k}l_i^2e^{-hl_i}-4\left(\sum_{i=1}^{k}l_ie^{-hl_i}\right)^2,$$
and

\begin{displaymath}
\left[\frac{\partial^3P(-hr+<u,f>)}{\partial \beta\partial u_i\partial u_j}\right]_{u=0}=
\left\{\begin{array}{ll}
-2l_ie^{-hl_i}+4e^{-hl_i}\sum_{s=1}^{k}l_se^{-hl_s} &\textrm{if $i=j$}\\
0 &\textrm {if $i\ne j$}.
\end{array} \right.
\end{displaymath}
These above imply that
\begin{eqnarray*}
\frac{\partial^4\beta (0)}{\partial u_i^4}&=&\frac{1}{2\sum_{s=1}^{k}l_se^{-hl_s}}\left\{2e^{-hl_i}-12e^{-2hl_i}+\left[2\sum_{s=1}^{k}l_s^2e^{-hl_s}-4\left(\sum_{s=1}^{k}l_se^{-hl_s}\right)^2\right]\right.\\
&\times&\frac{3e^{-2hl_i}}{\left(\sum_{s=1}^{k}l_se^{-hls}\right)^2}
+\left.6\left(-2l_ie^{-hl_i}+4e^{-hl_i}\sum_{s=1}^{k}l_se^{-hl_s}\right)\cdot\frac{e^{-hl_i}}{\sum_{s=1}^{k}l_se^{-hl_s}}\right\}\\
&=&:\frac{8d_ie^{-2hl_i}}{\sum_{s=1}^{k}l_se^{-hl_s}},
\end{eqnarray*}
where
\begin{equation}
\label{di}
d_i=\frac{1}{16}\left\{2e^{hl_i}-\frac{12l_i}{\sum_{s=1}^{k}l_se^{-hl_s}}+\frac{6\sum_{s=1}^{k}l_s^2e^{-hl_s}}{(\sum_{s=1}^{k}l_se^{-hl_s}
)^2}\right\}.
\end{equation}
And
\begin{eqnarray*}
\frac{\partial^4\beta (0)}{\partial u_i^2\partial u_j^2}&=&\frac{1}{2\sum_{s=1}^{k}l_se^{-hl_s}}\left\{-4e^{-h(l_i+l_j)}+\left[2\sum_{s=1}^{k}l_s^2e^{-hl_s}-4\left(\sum_{s=1}^{k}l_se^{-hl_s}\right)^2\right]\right.\\
&\times & \frac{e^{-h(l_i+l_j)}}{\left(\sum_{s=1}^{k}l_se^{-hl_s}\right)^2}
+\left(-2l_ie^{-hl_i}+4e^{-hl_i}\sum_{s=1}^{k}l_se^{-hl_s}\right)\cdot\frac{e^{-hl_j}}{\sum_{s=1}^{k}l_se^{-hl_s}}\\
&+&\left.\left(-2l_je^{-hl_j}+4e^{-hl_j}\sum_{s=1}^{k}l_se^{-hl_s}\right)\cdot\frac{e^{-hl_i}}{\sum_{s=1}^{k}l_se^{-hl_s}}\right\}\\
&=&: \frac{24d_{ij}}{\sum_{s=1}^{k}l_se^{-hl_s}}e^{-h(l_i+l_j)},
\end{eqnarray*}
where
\begin{equation}
\label{dij}
d_{ij}=\frac{1}{24}\left\{\frac{\sum_{s=1}^{k}l_s^2e^{-hl_s}}{\left(\sum_{s=1}^{k}l_se^{-hl_s}
\right)^2}-\frac{l_i+l_j}{\sum_{s=1}^{k}l_se^{-hl_s}}\right\}.
\end{equation}
Otherwise,
$$\frac{\partial^4\beta (0)}{\partial u_i\partial u_j\partial u_m \partial u_n}=0.$$
Let $\sum_{i=1}^{k}l_ie^{-hl_i}=\frac{1}{c'}$. By the preceding
section we have
$$a_{ij}=2\pi^2\hat{g}(-ih)\int_{\mathbb{R}^k}e^{-\frac{1}{2}c'\sum_{m=1}^{k}e^{-hl_m}v_m^2}v_iv_jdv.$$
So if $i\ne j$, $a_{ij}=0$. For $i=j$,
\begin{eqnarray*}
a_{ii}&=&2\pi^2\hat{g}(-ih)\int_{\mathbb{R}^k}e^{-\frac{1}{2}c'\sum_{m=1}^{k}e^{-hl_m}v_m^2}v_i^2dv\\
&=&2\pi^2\hat{g}(-ih)\frac{\sqrt{2\pi}}{\sqrt{c'e^{-hl_1}}}\times\cdots\times\frac{e^{hl_i}\sqrt{2\pi}}{c'\sqrt{c'e^{-hl_i}}}\times\cdots\times\frac{\sqrt{2\pi}}{\sqrt{c'e^{-hl_k}}}\\
&=& \frac{(2\pi)^{\frac{k}{2}+2}\hat{g}(-ih)e^{hl_i}}{2{c'}^{\frac{k}{2}+1}\sqrt{e^{-h(l_1+\cdots+l_k)}}}.
\end{eqnarray*}
Substituting $\hat{g}(-ih)=1/h$ and let $\xi=\frac{(2\pi)^{\frac{k}{2}+2}}{2h{c'}^{\frac{k}{2}+1}\sqrt{e^{-h(l_1+\cdots+l_k)}}}$,
\begin{displaymath}
a_{ij}=\left\{ \begin{array}{ll}
\xi e^{hl_i}& \textrm{if $i=j$}\\
0&\textrm{if $i\ne j$}.
\end{array}\right.
\end{displaymath}
In order to obtain constant $c_{1,0}$, we first calculate $F_2(iv)$. Since
$$\beta^{(4)}(0)\cdot (iv)^4=24c'\sum_{i\ne j}^{}d_{ij}e^{-h(l_i+l_j)}(iv_i)^2(iv_j)^2+8c'\sum_{i=1}^{k}d_ie^{-2hl_i}(iv_i)^4,$$
we have
\begin{eqnarray*}
&&F_2(iv)= \frac{3\hat{g}(-ih)}{72}\\
&&\times\left[24c'\sum_{i\ne j}^{}d_{ij}e^{-h(l_i+l_j)}(iv_i)^2(iv_j)^2+8c'\sum_{i=1}^{k}d_ie^{-2hl_i}(iv_i)^4\right]
+\frac{\bar{g}^{(2)}_0(0)}{2}\cdot (iv)^2+\bar{g}^{(0)}_1(0)\\
&&= c'\hat{g}(-ih)\sum_{i\ne j}^{}d_{ij}e^{-h(l_i+l_j)}v_i^2v_j^2+\frac{c'\hat{g}(-ih)}{3}\sum_{i=1}^{k}d_ie^{-2hl_i}v_i^4
+\frac{\bar{g}^{(2)}_0(0)}{2}\cdot (iv)^2+\bar{g}^{(0)}_1(0).
\end{eqnarray*}
Hence
\begin{eqnarray*}
&&c_{1,0}=\int_{\mathbb{R}^k}e^{-\frac{1}{2}c'\sum_{i=1}^{k}e^{-hl_i}v_i^2}F_2(iv)dv
= c'\hat{g}(-ih)\sum_{i\ne j}^{}d_{ij}e^{-h(l_i+l_j)}\\
&&\times\frac{\sqrt{2\pi}}{\sqrt{c'e^{-hl_1}}}\times\cdots\frac{e^{hl_i}\sqrt{2\pi}}{c'\sqrt{c'e^{-hl_i}}}\times\cdots\times\frac{e^{hl_j}\sqrt{2\pi}}{c'\sqrt{c'e^{-hl_j}}}\times\cdots\times\frac{\sqrt{\pi}}{\sqrt{c'e^{-hl_k}}}\\
&&+\frac{c'\hat{g}(-ih)}{3}\sum_{i=1}^{k}d_ie^{-2hl_i}\frac{\sqrt{2\pi}}{\sqrt{c'e^{-hl_1}}}
\times\cdots\frac{3e^{2hl_i}\sqrt{2\pi}}{{c'}^2\sqrt{c'e^{-hl_i}}}\times\cdots\times\frac{\sqrt{2\pi}}{\sqrt{c'e^{-hl_k}}}
\end{eqnarray*}
\begin{eqnarray*}
&&+\int_{\mathbb{R}^k}e^{-\frac{1}{2}c'\sum_{i=1}^{k}e^{-hl_i}v_i^2}\left(\frac{\bar{g}^{(2)}_0(0)}{2}\cdot
(iv)^2+\bar{g}^{(0)}_1(0)\right)dv\\
&&= d_1\hat{g}(-ih)\frac{(2\pi)^{\frac{k}{2}}}{{c'}^{\frac{k}{2}+1}\sqrt{e^{-h(l_1+\cdots+l_k)}}}+d_2\hat{g}(-ih)\frac{(2\pi)^{\frac{k}{2}}}{{c'}^{\frac{k}{2}+1}\sqrt{e^{-h(l_1+\cdots+l_k)}}}+C\\
&&=\frac{1}{2\pi^2}(d_1+d_2)\xi+C,
\end{eqnarray*}
where
\begin{equation}
\label{d1}
d_1=\sum_{i\ne j}^{}d_{ij} \qquad d_2=\sum_{i=1}^{k}d_i
\end{equation}
 and
\begin{eqnarray*}
C&=&\int_{\mathbb{R}^k}e^{-\frac{1}{2}c'\sum_{i=1}^{k}e^{-hl_i}v_i^2}\left(\frac{\bar{g}^{(2)}_0(0)}{2}\cdot (iv)^2+\bar{g}^{(0)}_1(0)\right)dv\\
&=& \int_{\mathbb{R}^k}e^{-\frac{1}{2}c'\sum_{i=1}^{k}e^{-hl_i}v_i^2}\left(-\frac{c'}{2h^2}\sum_{i=1}^{k}e^{-hl_i}v_i^2-\frac{1}{h^2}\right)dv\\
&=&-\frac{k}{2h^2}\frac{(2\pi)^{\frac{k}{2}}}{{c'}^{\frac{k}{2}}\sqrt{e^{-h(l_1+\cdots+l_k)}}}-\frac{1}{h^2}\frac{(2\pi)^{\frac{k}{2}}}{{c'}^{\frac{k}{2}}\sqrt{e^{-h(l_1+\cdots+l_k)}}}\\
&=&-\frac{k+2}{4h\pi^2}c'\xi.
\end{eqnarray*}
So we have
\newtheorem{th43}[th41]{Theorem}
\begin{th43} For $G$ is a graph with one vertex and $k$ edges which form $k$ loops, we have
$$\pi(T,\alpha)=\frac{e^{Th}}{T^{b/2+1}}\left(c_0+\sum_{n=1}^N
 \frac{c_{n}(\alpha)}{T^{n}}+O\left(\frac{1}{T^{N+1}}\right)\right)
 \textrm{ as } T\to \infty. $$
The first error term $c_1(\alpha)$ is following.
$$c_{1}(\alpha)=-\sum_{i=1}^{k}\xi e^{hl_i}\alpha_i^2+c_{1,0},$$
where
$\xi=\frac{1}{2h}(2\pi)^{\frac{k}{2}+2}\sqrt{e^{h(l_1+l_2+\cdots+l_k)}}\left(\sum_{i=1}^{k}l_ie^{-hl_i}\right)^{\frac{k}{2}+1}$ and\\
$c_{1,0}=\frac{1}{2\pi}\left(d_1+d_2-\frac{k+2}{2h\sum_{i=1}^{k}l_ie^{-hl_i}}\right)\xi$,
$d_1$ and $d_2$ are specified by (\ref{di}), (\ref{dij}) and (\ref{d1}).
\end{th43}
\vskip.3cm
\noindent Especially, if $k=2$, $\alpha=(\alpha_1,\alpha_2)\in\mathbb{Z}^2$ then

$$c_1(\alpha)=-\frac{4\pi^3(l_1e^{-hl_1}+l_2e^{-hl_2})^2\sqrt{e^{h(l_1+l_2)}}}{h}\left(e^{hl_1}\alpha_1^2+e^{hl_2}\alpha_2^2\right)+c_{1,0}.$$
Since $h$ satisfies $e^{-hl_1}+e^{-hl_2}=\frac{1}{2}$, the constant $c_{1,0}$ is
\begin{eqnarray*}
c_{1,0}&=&\frac{4\pi^3\sqrt{e^{h(l_1+l_2)}}}{96h}\left\{[108+12(e^{hl_1}+e^{hl_2})](l_1e^{-hl_1}+l_2e^{-hl_2})^2\right.\\
&&-\left.38l_1l_2-63(l_1^2e^{-hl_1}+l_2^2e^{-hl_2}) \right\}.
\end{eqnarray*}
\vskip.6cm

{\bf 6. Example 2.} \  Let $G$ be a graph with two vertices and
three edges which form two loops (Figure $6.1$). It can be coded
with the following directed graph (Figure $6.2$).
\begin{figure}[!ht]
\label{fig1.2}
\begin{center}
\includegraphics[width=8cm]{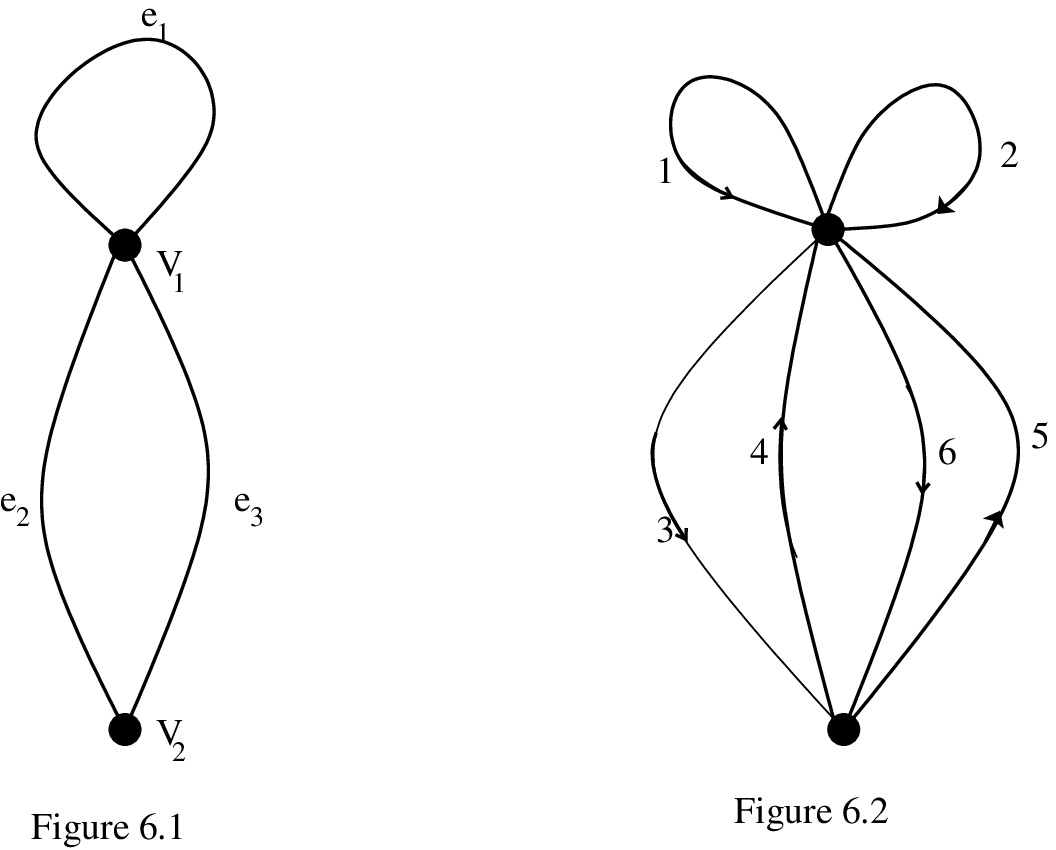}
\end{center}
\end{figure}

\vskip.5cm The matrix $A_G$ associated with $G_o$ (Figure $6.2$)
is
\begin{displaymath}
A_G= \left(\begin{array}{cccccc}
1 & 1 & 1 & 0 & 0 &1\\
1 &1 &1& 0& 0& 1\\
0&0&0&1&1&0\\
1&1&1&0&0&1\\
1&1&1&0&0&1\\
0&0&0&1&1&0\\
\end{array}
\right)
\end{displaymath}
Let the lengths of $e_1, e_2, e_3$ be $l_1, l_2$ and $ l_3$, respectively such that conditions $(A)$ and $(B)$ satisfied.

We define
\begin{displaymath}
r(x)=r(x_0)=\left\{\begin{array}{ll}
l_1&\textrm{if $x_0=1$ or $x_0=2$}\\
l_2 &\textrm{if $x_0=3$ or $x_0=4$}\\
l_3 &\textrm{if $x_0=5$ or $x_0=6$}.
\end{array} \right.
\end{displaymath}
And $f(x)=f(x_0)=(f_1(x_0),f_2(x_0))$ such that
\begin{displaymath}
f_1(x)=f_1(x_0)=\left\{\begin{array}{ll}
1&\textrm{if $x_0=1$}\\
-1 &\textrm{if $x_0=2$}\\
0 &\textrm{otherwise},
\end{array}\right.
\end{displaymath}
and
\begin{displaymath}
f_2(x)=f_2(x_0)=\left\{\begin{array}{ll}
\frac{1}{2}&\textrm{if $x_0=3$ or $x_0=5$}\\
-\frac{1}{2} &\textrm{if $x_0=4$ or $x_0=6$}\\
0 &\textrm{otherwise}.
\end{array}\right.
\end{displaymath}
In this case, $H_1(G,\mathbb{Z})=\mathbb{Z}^2$. there exist a measure $\mu_{-hr}$ which is Markov measure. That is if we denote by $\mu_{-hr}(x_0,x_1,\ldots,x_n)$ the measure of cylinder $\{x:x=x_0x_1\ldots x_n *\cdots\}$ then
\begin{eqnarray*}
&(1)& \mu_{-hr}(x_0,x_1,\ldots,x_n)\ge 0;\\
&(2)& \sum_{x_0}^{}\mu_{-hr}(x_0)=1;\\
&(3)& \mu_{-hr}(x_0,x_1,\ldots,x_n)=\sum_{x_{n+1}}^{}\mu_{-hr}(x_0,x_1,\ldots,x_{n+1}).
\end{eqnarray*}
In this case, it is also satisfy following.
$$\mu_{-hr}(1)=\mu_{-hr}(2),\quad \mu_{-hr}(3)=\mu_{-hr}(4),\quad \mu_{-hr}(5)=\mu_{-hr}(6).$$
If we write $n(1)=2$, $n(2)=1$, $n(3)=4$, $n(4)=3$, $n(5)=6$ and $n(6)=5$,  by the symmetry of graph $G$, we have
$$\mu_{-hr}(x_0, x_1, \ldots, x_{n-1})=\mu_{-hr}(n(x_{n-1}), \ldots, n(x_1), n(x_0)).$$

In order to calculate $\nabla ^2 \beta (0)$, we will use another expression for $\nabla ^2 \beta (0)$ in the form,
$$\frac{\partial^2\beta (0)}{\partial u_i \partial u_j}=\frac{1}{\int rd\mu_{-hr}}\lim_{n\to\infty}^{}\frac{1}{n}\int f_i^nf_j^nd\mu_{-hr}.$$
We first prove the following lemma by induction.
\newtheorem {le41}[le40]{Lemma}
\begin{le41}
$\forall n\in \mathbb{N}$,
$$\int f_1^nf_2^nd\mu_{-hr}=0$$
\end{le41}
\noindent\emph{Proof.} \\
(1) Since
$$f_1(x)f_2(x)\equiv 0,$$
by definition of $f$, lemma holds for $n=1$.\\
(2)We assume that lemma holds for $n=k\in \mathbb{N}$, then
$$\int f_1^kf_2^kd\mu_{-hr}=0.$$
That is
$$\sum_{x_0,x_1,\ldots,x_{k-1}}^{}\left(f_1(x_0)+\cdots+f_1(x_{k-1})\right)\left(f_2(x_0)+\cdots+f_2(x_{k-1})\right)\mu_{-hr}(x_0,\ldots,x_{k-1})\\=0.$$
For $n=k+1$,
\begin{eqnarray*}
&&\int f_1^{k+1}f_2^{k+1}d\mu_{-hr}=\sum_{x_0,\ldots,x_{k-1},x_k}^{}\left(f_1(x_0)+\cdots+f_1(x_{k-1})+f_1(x_k)\right)\\
&\times&\left(f_2(x_0)+\cdots+f_2(x_{k-1})+f_2(x_k)\right)\mu_{-hr}(x_0,\ldots,x_{k-1},x_k)\\
&=&\sum_{x_0,\ldots,x_{k-1},x_k}^{}\left(f_1(x_0)+\cdots+f_1(x_{k-1})\right)\left(f_2(x_0)+\cdots+f_2(x_{k-1})\right)\mu_{-hr}(x_0,\ldots,x_{k-1},x_k)\\
&+&\sum_{x_0,\ldots,x_{k-1},x_k}^{}f_1(x_k)\left(f_2(x_0)+\cdots+f_2(x_{k-1})\right)\mu_{-hr}(x_0,\ldots,x_{k-1},x_k)
\end{eqnarray*}
\begin{eqnarray*}
&+&\sum_{x_0,\ldots,x_{k-1},x_k}^{}f_2(x_k)\left(f_1(x_0)+\cdots+f_1(x_{k-1})\right)\mu_{-hr}(x_0,\ldots,x_{k-1},x_k)\\
&+&\sum_{x_0,\ldots,x_{k-1},x_k}^{}f_1(x_k)f_2(x_k)\mu_{-hr}(x_0,\ldots,x_{k-1},x_k).
\end{eqnarray*}
By induction assumption, we have
\begin{eqnarray*}
&&\sum_{x_0,\ldots,x_{k-1},x_k}^{}\left(f_1(x_0)+\cdots+f_1(x_{k-1})\right)\left(f_2(x_0)+\cdots+f_2(x_{k-1})\right)\mu_{-hr}(x_0,\ldots,x_{k-1},x_k)\\
&=&\sum_{x_0,\ldots,x_{k-1}}^{}\left(f_1(x_0)+\cdots+f_1(x_{k-1})\right)\left(f_2(x_0)+\cdots+f_2(x_{k-1})\right)\sum_{x_k}^{}\mu_{-hr}(x_0,\ldots,x_{k-1},x_k)\\
&=&\sum_{x_0,\ldots,x_{k-1}}^{}\left(f_1(x_0)+\cdots+f_1(x_{k-1})\right)\left(f_2(x_0)+\cdots+f_2(x_{k-1})\right)\mu_{-hr}(x_0,\ldots,x_{k-1})\\
&=&\int f_1^k f_2^kd\mu_{-hr}=0.
\end{eqnarray*}
Since
$$A(x_{k-1},1)=1\Longleftrightarrow A(x_{k-1},2)=1,$$
and
$$\mu_{-hr}(x_0, \ldots, x_{k-1}, 1)=\mu_{-hr}(x_0, \ldots, x_{k-1}, 2),$$
we have
\begin{eqnarray*}
&&\sum_{x_0,\ldots,x_{k-1},x_k}^{}f_1(x_k)\left(f_2(x_0)+\cdots+f_2(x_{k-1})\right)\mu_{-hr}(x_0,\ldots,x_{k-1},x_k)\\
&=&\sum_{x_0,\ldots,x_{k-1},1}^{}f_1(1)\left(f_2(x_0)+\cdots+f_2(x_{k-1})\right)\mu_{-hr}(x_0,\ldots,x_{k-1},1)\\
&+&\sum_{x_0,\ldots,x_{k-1},2}^{}f_1(2)\left(f_2(x_0)+\cdots+f_2(x_{k-1})\right)\mu_{-hr}(x_0,\ldots,x_{k-1},2)\\
&=&0.
\end{eqnarray*}
Similarly,
$$\sum_{x_0,\ldots,x_{k-1},x_k}^{}f_2(x_k)\left(f_1(x_0)+\cdots+f_1(x_{k-1})\right)\mu_{-hr}(x_0,\ldots,x_{k-1},x_k)=0.$$
It is always true for
$$\sum_{x_0,\ldots,x_{k-1},x_k}^{}f_1(x_k)f_2(x_k)\mu_{-hr}(x_0,\ldots,x_{k-1},x_k)=0.$$
So
$$\int f_1^{k+1}f_2^{k+1}d\mu_{-hr}=0.$$
\noindent (3) Hence
$$\forall n\in\mathbb{N}, \int f_1^nf_2^nd\mu_{-hr}=0.$$

\noindent Similarly, $$\forall n\in\mathbb{N}, \int f_2^nf_1^nd\mu_{-hr}=0.$$
We also need to calculate $\int (f_1^n)^2 d\mu_{-hr}$ and $\int (f_2^n)^2d\mu_{-hr}$. We have
\newtheorem{le42}[le40]{Lemma}
\begin{le42}
$\forall n\in\mathbb{N}$,
$$\int (f_1^n)^2 d\mu_{-hr}=n(\mu_{-hr}(1)+\mu_{-hr}(2))=2n\mu_{-hr}(1).$$
And
$$\int (f_2^n)^2 d\mu_{-hr}=\frac{1}{4}n(\mu_{-hr}(3)+\mu_{-hr}(4)+\mu_{-hr}(5)+\mu_{-hr}(6))=\frac{1}{2}n(\mu_{-hr}(3)+\mu_{-hr}(5)).$$
\end{le42}
\noindent\emph{Proof.}
 We prove this Lemma by induction as we did for Lemma $2$.
\vskip.2cm
\noindent Hence we have
\begin{displaymath}
\mathbf{\beta ''(0)}=\frac{1}{\int rd\mu_{-hr}}
\left(\begin{array}{cc}
2\mu_{-hr}(1)&0\\
0&\frac{1}{2}(\mu_{-hr}(3)+\mu_{-hr}(5)).\\
\end{array}
\right)
\end{displaymath}
As we see that in last section,
$$\mu_{-hr}(1)=\mu_{-hr}(2)=e^{-hl_1},$$
and
$$\mu_{-hr}(3)=\mu_{-hr}(4)=e^{-hl_2},$$
and
$$\mu_{-hr}(5)=\mu_{-hr}(6)=e^{-hl_3}.
$$
So
\begin{displaymath}
\mathbf{\beta ''(0)}=\frac{1}{2(l_1e^{-hl_1}+l_2e^{-hl_2}+l_3e^{-hl_3})}
\left(\begin{array}{cc}
2e^{-hl_1}&0\\
0&\frac{1}{2}(e^{-hl_2}+e^{-hl_3}).\\
\end{array}
\right)
\end{displaymath}
Now we can obtain $a_{ij}$. Since $\beta''(0)$ is diagonal, we still have $a_{12}=a_{21}=0$. Let
$$c'=\frac{1}{\int rd\mu_{-hr}}=\frac{1}{2(l_1e^{-hl_1}+l_2e^{-hl_2}+l_3e^{-hl_3})},$$ then
\begin{eqnarray*}
a_{11}&=& \frac{2\pi^2}{h}\int_{-\infty}^{+\infty}\int_{-\infty}^{+\infty}e^{-\frac{1}{2}c'(2e^{-hl_1}v_1^2+\frac{1}{2}(e^{-hl_2}+e^{-hl_3})v_2^2)}v_1^2dv_1dv_2\\
&=& \frac{2\pi^2}{h}\frac{\sqrt{2\pi}}{\sqrt{\frac{1}{2}c'(e^{-hl_2}+e^{-hl_3})}}\frac{\sqrt{2\pi}}{2c'e^{-hl_1}\sqrt{2c'e^{-hl_1}}}\\
&=&\frac{2\pi^3}{{c'}^2h}\frac{e^{hl_1}}{\sqrt{e^{-h(l_1+l_2)}+e^{-h(l_1+l_3)}}}\end{eqnarray*}
and
\begin{eqnarray*}
a_{22}&=& \frac{2\pi^2}{h}\int_{-\infty}^{+\infty}\int_{-\infty}^{+\infty}e^{-\frac{1}{2}c'(2e^{-hl_1}v_1^2+\frac{1}{2}(e^{-hl_2}+e^{-hl_3})v_2^2)}v_2^2dv_1dv_2\\
&=& \frac{2\pi^2}{h}\frac{\sqrt{2\pi}}{\frac{1}{2}c'(e^{-hl_2}+e^{-hl_3})\sqrt{\frac{1}{2}c'(e^{-hl_2}+e^{-hl_3})}}\frac{\sqrt{2\pi}}{\sqrt{2c'e^{-hl_1}}}\\
&=&\frac{8\pi^3}{{c'}^2h}\frac{e^{h(l_2+l_3)}}{(e^{hl_2}+e^{hl_3})\sqrt{e^{-h(l_1+l_2)}+e^{-h(l_1+l_3)}}}.
\end{eqnarray*}
Let
$$c=\frac{8\pi^3 (l_1e^{-hl_1}+l_2e^{-hl_2}+l_3e^{-hl_3})^2}{h\sqrt{e^{-h(l_1+l_2)}+e^{-h(l_1+l_3)}}}.$$
We have
\newtheorem{th44}[th41]{Theorem}
\begin{th44} Let $G$ be a graph with two vertices and three edges which form two loops.
$$\pi(T,\alpha)=\frac{e^{Th}}{T^{b/2+1}}\left(c_0+\sum_{n=1}^N \frac{c_{n}(\alpha)}{T^{n}}+O\left(\frac{1}{T^{\delta}}\right)\right) \textrm{ as } T\to \infty $$ with
$$c_{1}(\alpha)=-ce^{hl_1}\alpha_1^2-4c\frac{e^{h(l_2+l_3)}}{e^{hl_2}+e^{hl_3}}\alpha_2^2 +c_{1,0},$$
where
$$c=\frac{8\pi^3 (l_1e^{-hl_1}+l_2e^{-hl_2}+l_3e^{-hl_3})^2}{h\sqrt{e^{-h(l_1+l_2)}+e^{-h(l_1+l_3)}}}.$$
 and $c_{1,0}$ is a constant (which , since it is rather  complicated, we do not specify  here).
\end{th44}
\noindent{ACKNOWLEDGEMENTS}.\quad The author would like to thank
Richard Sharp and Mark Pollicott for encouragement and fruitful
discussions and is grateful to  CVCP and The  University of
Manchester for financial support. \vskip.8cm \centerline{\large
REFERENCES} \vskip.4cm
{\small [$\boldsymbol{1}$]\quad N. Anantharaman. Precise counting
results for closed orbits of Anosov  flows. \emph{Ann. Sci.
E\'cole Norm. Sup}. (4) $\boldsymbol{33}$ (2000), 33-56.

 [$\boldsymbol{2}$]\quad R. Bowen. Symbolic dynamics
for hyperbolic flows. \emph{Amer. J. Math.} $\boldsymbol{95}$
(1973), 429-459.

[$\boldsymbol{3}$]\quad D. Dolgopyat. Prevalence of rapid mixing
in hyperbolic flows. \emph{Ergod. Th. \& Dynam. Sys.}
$\boldsymbol{18}$ (1998), 1097-1114.

[$\boldsymbol{4}$]\quad A. Katsuda \& T. Sunada. Closed orbit in
homology class. \emph{Inst. Hautes \'Etudes Sci. Publ. Math.}
$\boldsymbol{71}$ (1990), 5-32.

 [$\boldsymbol{5}$]\quad D. Liu.  Asymptotic expansion
for closed orbits in homology classes for Anosov flows, \emph{
 Math. Proc. Camb. Phil. Soc.} $\boldsymbol{136}$ (2004), 383-397.

[$\boldsymbol{6}$]\quad D. Liu. Asymptotic expansion for closed
geodesics in homology classes, \emph{ Glasgow Math. J.}
$\boldsymbol{46}$ (2004), 283-299.

 [$\boldsymbol{7}$]\quad A. Manning. Axiom A
diffeomorphisms have rational zeta functions. \emph{ Bull. London
Math. Soc.} $\boldsymbol{3}$ (1971), 215-220.

 [$\boldsymbol{8}$]\quad W. Parry \& M.
Pollicott. Zeta functions and the periodic orbit structure of
hyperbolic dynamics. \emph{Ast\'erisque} 187-188 (1990).

[$\boldsymbol{9}$]\quad M. Pollicott and R. Sharp. Asymptotic
expansions for closed orbits in homology classes. \emph{
Geometriae Dedicata}. $\boldsymbol{87}$
 (2001), 123-160.

[$\boldsymbol{10}$]\quad M. Pollicott and R. Sharp. Error terms
for closed orbits of hyperbolic flows. \emph{ Ergod. Th. \& Dynam.
Sys.} $\boldsymbol{21}$ (2001), 545-562.

[$\boldsymbol{11}$]\quad D. Ruelle. An extension of the theory of
Fredholm determinants. \emph{Inst. Hautes \'Etudes Sci. Publ.
Math}. $\boldsymbol{72}$ (1990), 175-193.

[$\boldsymbol{12}$]\quad R. Sharp.  Closed orbits in homology
classes for Anosov flows. \emph{Ergod. Th. \& Dynam. Sys.}
$\boldsymbol{13}$ (1993), 387-408.}

\end{document}